\documentclass{article}

\usepackage{ifthen} 
\usepackage{latexsym} 
\usepackage{substr} 

\usepackage{graphics}
\usepackage{graphicx}

\newcommand{\NC}{\newcommand}
\newcommand{\RC}{\renewcommand}

\NC{\ld}{\left} 
\NC{\rd}{\right} 

\NC{\ul}[1]{\underline{#1}} 
\NC{\ut}[1]{\underline{\underline{#1}}} 
\NC{\ct}{\ldots} 
\NC{\tr}[1]{\mathrm{#1}} 

\NC{\oo}{\infty} 
\NC{\ip}{\cdot} 
\NC{\op}{\times} 
\NC{\nd}{\frac} 
\NC{\ml}{\times} 
\NC{\mv}{\textrm{Max}} 
\NC{\re}{\textrm{Re}} 
\NC{\im}{\textrm{Im}} 
\NC{\sr}{\sqrt} 
\NC{\su}{\vert} 
\NC{\gt}{\rightarrow} 
\NC{\cv}{\ast} 
\RC{\ss}{\ell} 

\NC{\csch}{\mathrm{csch}} 
\NC{\sech}{\mathrm{sech}} 

\NC{\av}[1]{\ld| #1 \rd|} 
\NC{\iv}[1]{\ld( #1 \rd)} 
\NC{\pd}[2]{\partial_{#2} #1} 
\NC{\ci}[2]{\ld[ #1, #2 \rd]} 
\NC{\oi}[2]{\ld( #1, #2 \rd)} 

\NC{\grad}{\nabla} 
\NC{\divg}{\nabla \ip} 
\NC{\curl}{\nabla \op} 

\NC{\DD}[1]{\delta \iv{#1}} 
\NC{\US}[1]{{\cal U} \iv{#1}} 

\NC{\vt}[1]{\ul{#1}} 
\NC{\dy}[1]{\ut{#1}} 
\NC{\mr}[1]{\ld[ #1 \rd]} 

\NC{\UV}[1]{\vt{u}_{#1}} 
\NC{\NV}{\vt{0}} 
\NC{\ND}{\dy{0}} 
\NC{\ID}{\dy{I}} 
\NC{\NM}{\mr{\dy{0}}} 
\NC{\IM}{\mr{\dy{I}}} 

\NC{\PV}{\vt{r}} 
\NC{\PT}{\iv{\PV, t}} 
\NC{\ZB}[1]{z_{#1}} 

\NC{\Co}{c_{0}} 
\NC{\Eo}{\epsilon_{0}} 
\NC{\Mo}{\mu_{0}} 
\NC{\No}{\eta_{0}} 

\NC{\EF}{\vt{E}} 
\NC{\BF}{\vt{B}} 
\NC{\DF}{\vt{D}} 
\NC{\HF}{\vt{H}} 
\NC{\PF}{\vt{P}} 
\NC{\SF}{\vt{S}} 
\NC{\Sz}{S_{z}} 

\NC{\FQ}{\omega} 
\NC{\WL}{\lambda_{0}} 
\NC{\AF}{\omega} 

\NC{\EE}{\dy{\epsilon}_{r}} 
\NC{\SD}[1]{\dy{\chi}_{#1}} 
\NC{\RD}[1]{\dy{S}_{#1}} 
\NC{\SDr}{\dy{\chi}_{ref}} 

\NC{\AR}{\alpha} 
\NC{\HP}[1]{\IfSubStringInString{c}{#1}{\Omega}{\IfSubStringInString{s}{#1}{\Gamma}{}}} 
\NC{\SH}{h} 

\NC{\LT}[1]{\chi_{#1}} 
\NC{\OS}[1]{p_{#1}} 
\NC{\NP}{p_{nl}} 
\NC{\AP}[1]{N_{#1}} 
\NC{\RW}[1]{\lambda_{#1}} 
\NC{\RF}[1]{\omega_{#1}} 

\NC{\Ec}{\hat{\epsilon}_{2}}
\NC{\Ed}{\hat{\epsilon}_{d}}

\NC{\Nc}{n_{2}}
\NC{\Nd}{n_{d}}
\NC{\Vpc}{v_{p2}}
\NC{\Vpd}{v_{pd}}
\NC{\Vgc}{v_{g2}}
\NC{\Vgd}{v_{gd}}

\NC{\FF}{\vt{F}} 
\NC{\IF}{\vt{Q}} 
\NC{\VD}{\dy{V}} 
\NC{\WD}{\dy{W}} 

\NC{\EV}{\mr{\EF}} 
\NC{\FV}{\mr{\FF}} 
\NC{\IV}{\mr{\IF}} 
\NC{\SM}[1]{\mr{\SD{#1}}} 
\NC{\RM}[1]{\mr{\RD{#1}}} 
\NC{\VM}{\mr{\VD}} 
\NC{\WM}{\mr{\WD}} 

\NC{\dt}{\Delta t} 
\NC{\dz}{\Delta z} 
\NC{\bt}{\beta} 
\NC{\NS}[1]{N_{#1}} 

\NC{\LTc}[1]{\chi_{#1}^{c}} 
\NC{\SMc}[1]{\mr{\SD{#1}^{c}}} 
\NC{\IVc}{\mr{\IF^{c}}} 
\NC{\WMc}{\mr{\WD^{c}}} 
\NC{\WMi}{\mr{\WD'}} 

\NC{\PE}{\psi} 
\NC{\TC}{\tau_{0}} 
\NC{\TD}{t_{d}} 
\NC{\UE}{U} 
\NC{\UT}{U_{t}} 

\NC{\PW}{\vt{\varphi}} 
\NC{\CF}{\omega_{car}} 
\NC{\CP}{\phi} 
\NC{\CW}{\lambda_{car}} 
\NC{\TE}{\ld( s \rd)} 
\NC{\TM}{\ld( p \rd)} 

\NC{\FC}{f} 
\NC{\ZR}{z_{r}} 
\NC{\CEN}[1]{\zeta_{#1}} 
\NC{\RMS}[1]{\sigma_{#1}} 
\NC{\MOM}[2]{{\cal M}_{#1}^{\ld( #2 \rd)}} 
\NC{\COR}{{\cal C}} 

\NC{\TP}{\tau_{p}} 
\NC{\TU}{\tau_{u}} 
\NC{\TX}{\tau_{c}} 

\NC{\SP}{c_{p}} 
\NC{\SU}{c_{u}} 
\NC{\SX}{c_{c}} 

\NC{\FW}{0.4} 

\title{Swamping of circular Bragg phenomenon and shaping of videopulses}
\author{
Joseph B. Geddes III\footnote{Corresponding author; e--mail: jbgeddes3@psu.edu; phone: +1 814 278 1235.}~ and Akhlesh Lakhtakia \\
CATMAS---Computational and Theoretical Materials Sciences Group \\
Department of Engineering Science and Mechanics \\
212 Earth and Engineering Sciences Building \\
The Pennsylvania State University, University Park, PA 16802--6812, USA}
\date{20 June 2006}

\begin{document}

\maketitle

\begin{abstract}
We studied the durations and average speeds of videopulses transmitted through chiral sculptured thin films using a finite--difference algorithm. The chiral STF was either linear or possessed an intensity--dependent permittivity. Videopulse durations tended to decrease with increasing carrier wavelength and to increase, for given carrier wavelength and polarization state, when the chiral STF was nonlinear with intensity--dependent permittivity. The rate of decrease in durations with wavelength was less than that for longer--duration pulses. The effect of the circular Bragg phenomenon was swamped by the wide bandwidth possessed by videopulses. Two measures of videopulse average speed did not change much with carrier wavelength, and decreased slightly, for a given carrier wavelength and polarization state, in the nonlinear case. The other measure of average speed tended to increase with increasing carrier wavelength, and decrease, for a given carrier wavelength and polarization state, in the nonlinear case. 
\end{abstract}

\bigskip



\section{Introduction} \label{S: Introduction}

When circularly polarized light is incident normally on a periodically nonhomogeneous, structurally chiral medium of sufficient thickness---such as a chiral sculptured thin film (STF)---the circular Bragg phenomenon may occur~\cite{A.Lakhtakia-2005(B)}. The part of the light spectrum that falls within a range known as the Bragg regime is largely reflected from the chiral STF provided that the circular polarization state matches the structural handedness of the medium. If the circular polarization state is opposite, little light is reflected.

Research has elucidated the time--domain mechanism of the circular Bragg phenomenon as pulse bleeding, a continuous transfer of energy from the refracted pulse within the chiral STF to the reflected pulse outside of it~\cite{J.B.GeddesIII-2006(C)}. Pulse bleeding has also been predicted for incident videopulses, which are about one or two optical cycles in duration~\cite{J.B.GeddesIII-2002(P)}. The purpose of this communication is to extend previous work on the quantification of the durations and average speeds of ultrashort optical pulses through chiral STFs to videopulses~\cite{J.B.GeddesIII-2006(P)}. 

We computed three measures each of the durations and average speeds of videopulses transmitted through linear and nonlinear chiral STFs. The videopulses had different carrier wavelengths and polarization states, and we identified trends in duration and average speed as those parameters were varied.

Section~\ref{S: Theory} gives the constitutive relations of the chiral STFs studied, and explains the measures of duration and average speed we used. Section~\ref{S: Results} contains explanations of the trends we identified.

\section{Theory} \label{S: Theory}

Let us now describe the constitutive relations of the chiral STFs we studied and provide definitions of the measures of videopulse duration and average speed we used.

\subsection{Constitutive relations}

The polarization $\PF$ of a spatially local, dielectric, structurally right--handed chiral STF occupying the slab region $\ci{\ZB{L}}{\ZB{R}}$ depends on the electric field $\EF$ as follows:
\begin{equation} \label{E: Causal response}
\PF \PT  = \Eo \int_{0}^{t} \SD{e} \iv{\PV, t'} \ip \ld( 1 + \NP \av{\EF \iv{\PV, t - t'}}^{2}\rd) \EF \iv{\PV, t - t'} dt' \, .
\end{equation}
Here $\Eo$ is the permittivity of free space, $t$ is the time, and $\PV$ is the position vector defined with respect to a cartesian coordinate system defined with unit vectors $\ld\{ \UV{x}, \UV{y}, \UV{z} \rd\}$. The nonlinearity parameter $\NP$ determines whether the chiral STF will obey the superposition principle; when $\NP = 0$ the film is linear, but it exhibits an intensity--dependent permittivity otherwise~\cite{V.C.Venugopal-1998(P), R.W.Hellwarth-1977(P)}.

The dielectric susceptibility dyadic $\SD{e} \iv{\PV, t}$, which is null--valued for $z \notin \ci{\ZB{L}}{\ZB{R}}$, can be factored for $z \in \ci{\ZB{L}}{\ZB{R}}$ as follows: 
\begin{equation}
\SD{e} \PT = \RD{z} \iv{z - \ZB{L}} \ip \RD{y} \iv{\AR} \ip \SDr \iv{t} \ip \RD{y}^{-1} \iv{\AR} \ip \RD{z}^{-1} \iv{z - \ZB{L}} \, . 
\end{equation}
Here, the helicoidal rotation dyadic
\begin{equation}
\RD{z} \iv{z}
= \cos \ld( \nd{\pi z}{\HP{c}} \rd) \ld( \UV{x} \UV{x} + \UV{y} \UV{y} \rd)
+ \sin \ld( \nd{\pi z}{\HP{c}} \rd) \ld( \UV{y} \UV{x} - \UV{x} \UV{y} \rd)
+ \UV{z} \UV{z}
\end{equation}
contains the structural half--period $\HP{c}$, the chiral STF taken to be structurally right--handed; and the tilt dyadic
\begin{equation}
\RD{y} \iv{\AR}
= \cos \AR \ld( \UV{x} \UV{x} + \UV{z} \UV{z} \rd)
+ \sin \AR \ld( \UV{z} \UV{x} - \UV{x} \UV{z} \rd)
+ \UV{y} \UV{y}
\end{equation}
is parameterized by the angle of rise $\AR$. The reference susceptibility dyadic $\SDr$ is defined as
\begin{equation}
\SDr \iv{t} 
= \LT{1} \iv{t} \UV{x} \UV{x} 
+ \LT{2} \iv{t} \UV{y} \UV{y} 
+ \LT{3} \iv{t} \UV{z} \UV{z} \, ,
\end{equation}
where
\begin{equation} \label{E: Lorentz model}
\LT{\ss} \iv{t} = \OS{\ss} \RF{\ss} \exp \iv{- \nd{\RF{\ss} t}{2 \pi \AP{\ss}}} \sin \iv{\RF{\ss} t} \US{t} \, , \quad \ss = 1, 2, 3 \, .
\end{equation}
In the foregoing equation the oscillator strengths $\OS{\ss}$ and the parameters $\RW{\ss}$ and $\AP{\ss}$ quantify the resonance wavelengths and absorption characteristics of the chiral STF; $\RF{\ss} = 2 \pi \Co / \RW{\ss}$ are the resonance angular frequencies; and $\US{t}$ is the unit step function. 

\subsection{Pulse durations and average speeds}

A pulsed plane wave, consisting of a circularly polarized carrier plane wave $\PW \iv{t}$ that is amplitude--modulated by a pulse envelope $\PE \iv{t}$, was launched from the plane $z = 0$ towards the chiral STF. The carrier plane wave was characterized by the carrier wavelength $\CW$ and whether it was left circularly polarized (LCP) or right circularly polarized (RCP). The pulse envelope was gaussian as per
\begin{equation} \label{E: Pulse envelope}
\PE \iv{t} = \sr{\nd{\No \UT}{\TC \sr{\pi}}} \, \exp \iv{- \nd{1}{2} \ld( \nd{t - \TD}{\TC} \rd)^{2}} \, , \quad t > 0 \, ,
\end{equation}
with $\PE \iv{t} = 0$ for $t \leq 0$. The pulse envelope was characterized by the time constant $\TC$ and the time delay $\TD$. The total energy per unit area $\UT$ was chosen so that 
\begin{equation}
\int_{-\oo}^{\oo} \Sz \iv{z, t} dt = \UT \, , 
\end{equation}
where $\Sz = \UV{z} \ip \ld( \EF \op \HF \rd)$ is the axial component of the instantaneous Poynting vector. 

The propagation of the pulse was computed with a finite--difference method~\cite{J.B.GeddesIII-2006(P)}. We considered three measures of pulse duration---the equivalent, root mean square (RMS), and correlation---and three measures of average speed---the peak, center--of--energy, and correlation.

To define these measures, we used the following notation. The $m^{\tr{th}}$ moment $\MOM{\FC, \xi}{m}$ of a scalar function $\FC \iv{\xi}$ with respect to variable $\xi$ on the interval $\ci{\xi_{a}}{\xi_{b}}$ is
\begin{equation}
\MOM{\FC, \xi}{m} \iv{\FC \iv{\xi}, \ci{\xi_{a}}{\xi_{b}}} = \int_{\xi_{a}}^{\xi_{b}} \xi^{m} \FC \iv{\xi} d\xi \, ,
\end{equation}
and the RMS deviation from the centroid
\begin{equation}
\CEN{\FC, \xi} \iv{\FC \iv{\xi}, \ci{\xi_{a}}{\xi_{b}}} = \nd{\MOM{\FC, \xi}{1} \iv{\FC \iv{\xi}, \ci{\xi_{a}}{\xi_{b}}}}{\MOM{\FC, \xi}{0} \iv{\FC \iv{\xi}, \ci{\xi_{a}}{\xi_{b}}}}
\end{equation}
is~\cite{R.N.Bracewell-2000(B)}
\begin{eqnarray}
\RMS{\FC, \xi} \iv{\FC \iv{\xi}, \ci{\xi_{a}}{\xi_{b}}} 
= \sr{\nd{\MOM{\FC, \xi}{2} \iv{\FC \iv{\xi}, \ci{\xi_{a}}{\xi_{b}}}}{\MOM{\FC, \xi}{0}\iv{\FC \iv{\xi}, \ci{\xi_{a}}{\xi_{b}}}} - \CEN{\FC, \xi}^{2} \iv{\FC \iv{\xi}, \ci{\xi_{a}}{\xi_{b}}}} \, .
\end{eqnarray}
When $\MOM{\FC, \xi}{0} \iv{\FC \iv{\xi}, \ci{\xi_{a}}{\xi_{b}}} = 0$, both $\CEN{\FC, \xi}$ and $\RMS{\FC, \xi}$ are undefined. The electromagnetic energy density is
\begin{equation}
\UE \iv{z, t} = \Eo \av{\EF \iv{z, t}}^2 + \Mo \av{\HF \iv{z, t}}^2 \, .
\end{equation}
We recorded $\UE \iv{\ZR, t}$ of the transmitted pulse over the interval $\ci{t_{a}}{t_{b}}$ at $\ZR > \ZB{R}$. The equivalent duration~\cite{R.N.Bracewell-2000(B)} is defined as
\begin{equation}
\TP = \nd{\MOM{\UE, t}{0} \iv{\UE \iv{\ZR, t}, \ci{t_{a}}{t_{b}}}}{\mv \iv{\UE \iv{\ZR, t}, \ci{t_{a}}{t_{b}}}} \, ,
\end{equation}
where $\mv$ indicates, within the chosen interval, the maximum value of a function. The RMS duration is defined as
\begin{equation}
\TU = 2 \RMS{\UE, t} \iv{\UE \iv{\ZR, t}, \ci{t_{a}}{t_{b}}} \, .
\end{equation}
The correlation duration $\TX$ is defined as~\cite{R.N.Bracewell-2000(B)}
\begin{equation}
\TX = \nd{\int_{-\oo}^{\oo} \int_{-\oo}^{\oo} \UE \iv{\ZR, t} \UE \iv{\ZR, t + t'} dt' dt}{\MOM{\UE^{2}, t}{0} \iv{\UE^{2} \iv{\ZR, t}, \iv{-\oo, \oo}}} \, .
\end{equation} 

We found the times
\begin{eqnarray}
t_{p} & = & t_{\textrm{Max}} \iv{\UE \iv{\ZR, t}, \ci{t_{a}}{t_{b}}} \, , \\
t_{u} & = & \CEN{\UE, t} \iv{\UE \iv{\ZR, t}, \ci{t_{a}}{t_{b}}} \, , \\
t_{c} & = & \TD + t_{\textrm{Max}} \iv{\int_{0}^{\oo} \UE \iv{0, t'} \UE \iv{\ZR, t + t'} dt'} \, ,
\end{eqnarray}
where the function $t_{\textrm{Max}}$ indicates the time at which the maximum value occurs. Then, the average peak speed $\SP$, center--of--energy speed $\SU$, and correlation speed $\SX$~\cite{S.C.Bloch-1977(P)} were calculated as follows:
\begin{equation}
\nd{c_{\ss}}{\Co} = \nd{\ZB{R} - \ZB{L}}{\Co \ld( t_{\ss} - \TD \rd) - \ld( z_{r} - \ld( \ZB{R} - \ZB{L} \rd) \rd)} \, , \quad \ss = p, u, c \, .
\end{equation}
In our calculations, we used the rectangular rule to approximate integrals~\cite{Y.Jaluria-1996(B)}.

\section{Results} \label{S: Results}

\begin{figure}
\begin{center}
\scalebox{\FW}{\includegraphics{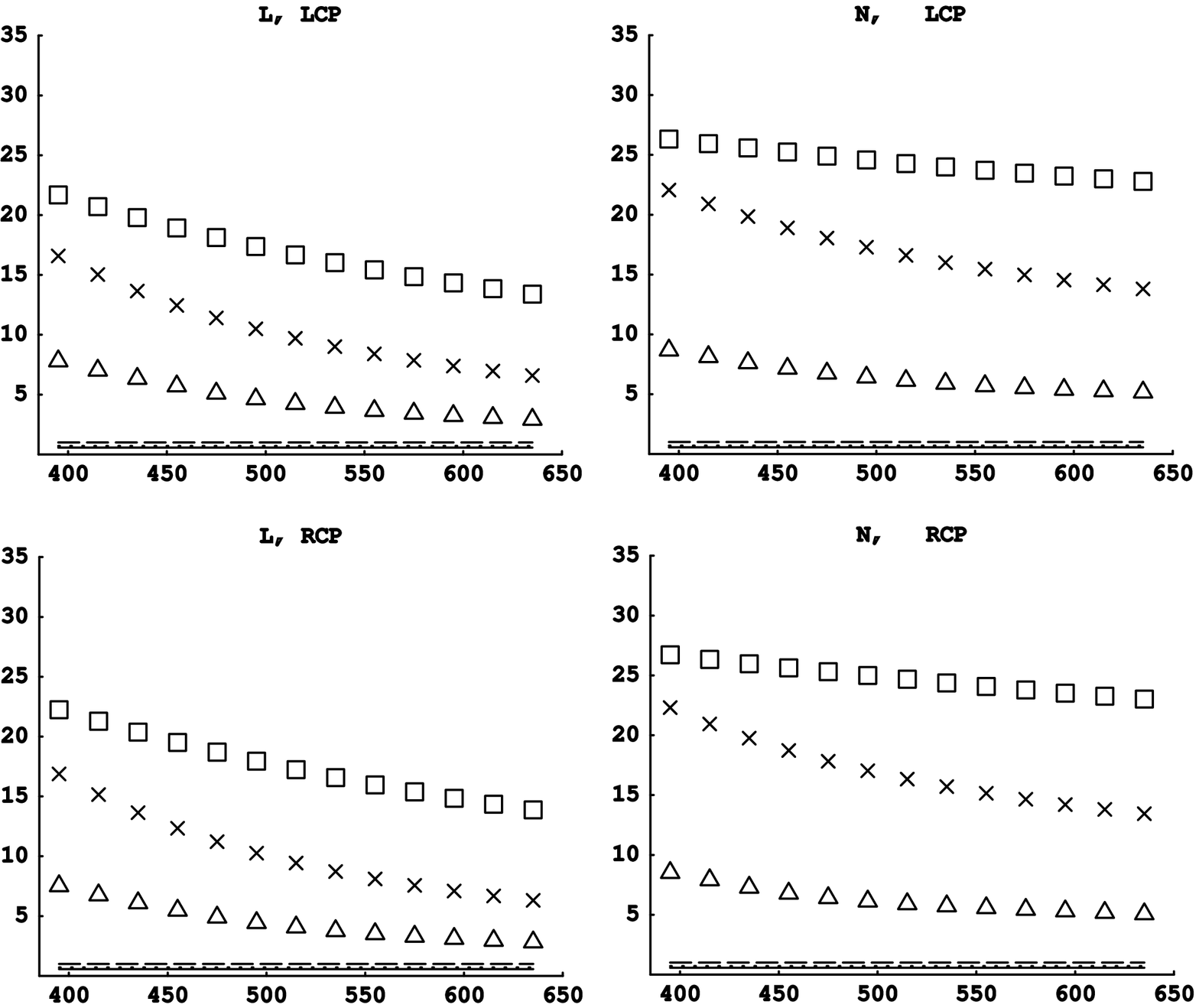}}
\end{center}
\caption{Durations $\TP$ ($\triangle$, in~fs), $\TU$ ($\Box$, in~fs), $\TX$ ($\ml$, in~fs) as functions of carrier wavelength $\CW$ (in~nm), as evaluated at $\ZR = 36$~$\mu$m over the interval $\ci{0}{190}$~fs. The carrier wavelength is either LCP (top) or RCP (bottom), and the chiral STF is either linear (case~L, $\NP = 0$, left) or nonlinear (case~N, $\NP = 3 \op 10^{-24}$~m$^2$/V$^{2}$, right). The lines at the bottom of each plot indicate the value of $\TP$ (dotted), $\TU$ (solid), and $\TX$ (dashed) for the incident pulses.} \label{F: Durations}
\end{figure}

\begin{figure}
\begin{center}
\scalebox{\FW}{\includegraphics{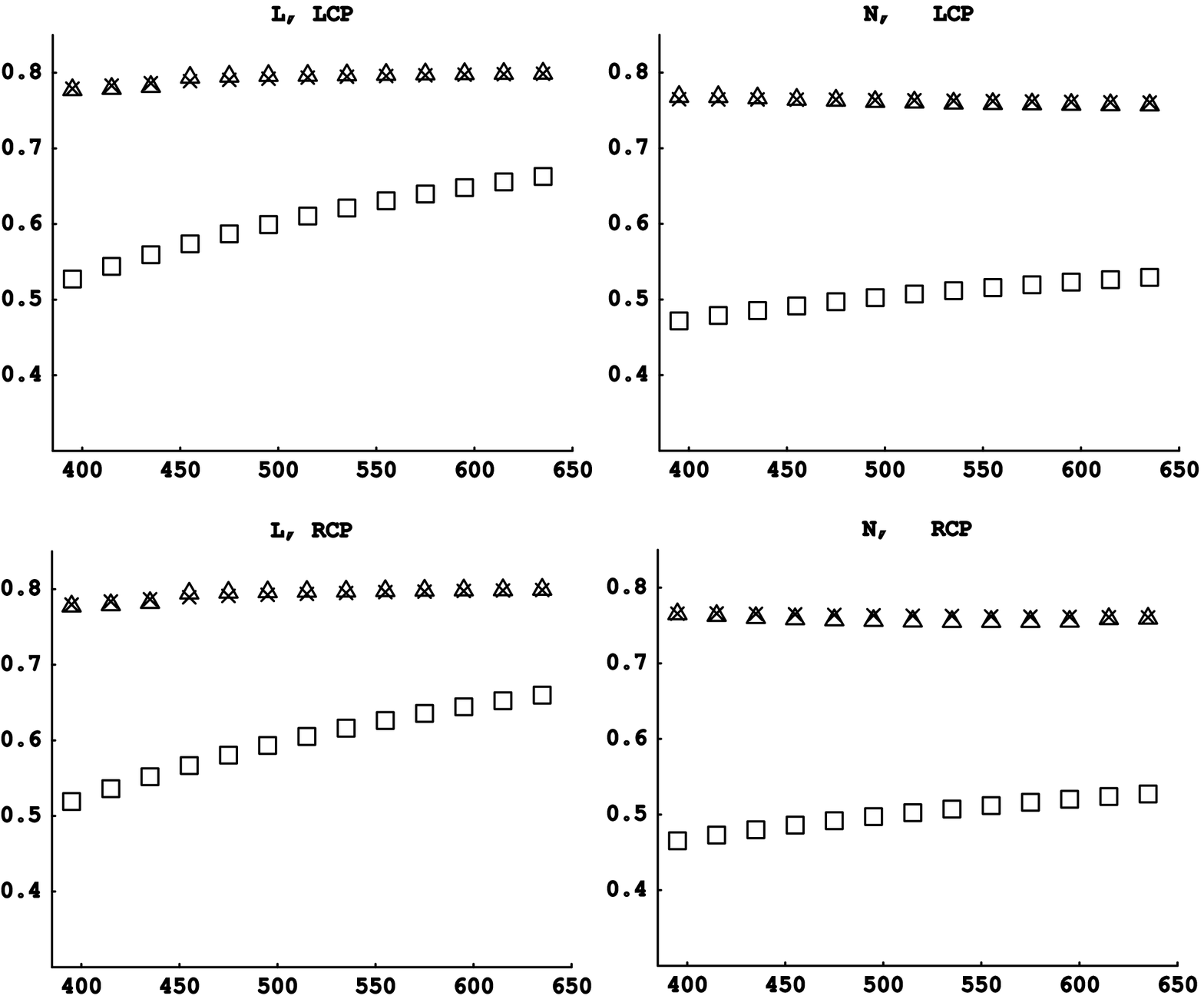}}
\end{center}
\caption{Normalized average speeds $\SP/\Co$ ($\triangle$), $\SU/\Co$ ($\Box$), and $\SX/\Co$ ($\ml$) as functions of carrier wavelength $\CW$ (in~nm), as evaluated at $\ZR = 36$~$\mu$m over the interval $\ci{0}{190}$~fs. The carrier wavelength is either LCP (top) or RCP (bottom), and the chiral STF is either linear (case~L, $\NP = 0$, left) or nonlinear (case~N, $\NP = 3 \op 10^{-24}$~m$^2$/V$^{2}$, right).} \label{F: Average speeds}
\end{figure}

The results of our calculations for incident videopulses are presented in Figures~\ref{F: Durations} and~\ref{F: Average speeds}. The parameters describing the chiral STF were chosen as follows: $\OS{1} = 0.52$, $\OS{2} = 0.42$, $\OS{3} = 0.40$, $\AP{1,2,3} = 100$, $\RW{1} = 290$~nm, $\RW{2,3} = 280$~nm, $\HP{c} = 200$~nm, and $\AR = 20^{\circ}$; the film was either linear (case~L, $p_{nl} = 0$) or nonlinear (case~N, $p_{nl} = 3 \op 10^{-24}$~m$^{2}$/V$^{2}$). The chiral STF was ten pitches thick ($\ZB{L} = 30$~$\mu$m and $\ZB{R} = 34$~$\mu$m). These parameters fix the center wavelength of the Bragg regime of the linear chiral STF to $\approx 516$~nm and the full--width half--maximum bandwidth to $\approx 27$~nm~\cite{J.B.GeddesIII-2001(P), A.Lakhtakia-1999(P)}; but both quantitites will shift slightly when $\NP = 3 \op 10^{-24}$~m$^{2}$/V$^{2}$~\cite{J.B.GeddesIII-2003(P)}. The incident videopulses had the following parameters: $\UT = 1 \op 10^{6}$~J/m$^{2}$, $\TC = 0.4$~fs, and $\TD = 1.6$~fs. They therefore comprised about $\nd{1}{2}$ optical cycle depending on the carrier wavelength, which was varied from 395--635~nm in increments of 20~nm.
 
Several general trends are evident in Figures~\ref{F: Durations} and~\ref{F: Average speeds}. The durations tended to decrease with increasing carrier wavelength, and tended to increase, for given carrier wavelength and polarization state, in going from the linear case to the nonlinear one. These two trends were also noted in an earlier study for longer--duration ($\TC = 2$~fs) incident pulses~\cite{J.B.GeddesIII-2006(P)}. There are several differences between these results and the results of the previous study. In Figure~\ref{F: Durations}, there is a large spread between the three measures of duration. Moreover, the decrease in durations with wavelength is less drastic in the case of videopulses than with the longer--duration pulses we studied earlier. This comparison can be explained by recourse to frequency--domain arguments as follows. The canonical refractive indexes of a chiral STF decrease with increasing free--space wavelength above the resonance wavelengths, along with the dispersion exhibited by the film~\cite{J.B.GeddesIII-2006(P)}. This decrease accounts for the decrease in durations with increasing wavelength, as there is less dispersion at higher wavelengths to elongate the pulse. As the duration of the incident pulse is decreased to that of a videopulse, its bandwidth widens accordingly.  This wider bandwidth causes the pulse to disperse more, for a given carrier wavelength, than a pulse with narrower bandwidth. Therefore, the durations of transmitted pulses decrease at a slower rate with increasing wavelength when the incident pulses are of shorter duration.

Moreover, in our earlier work, we found that that there is a local maximum in the RMS measure of duration when the chiral STF is linear and the circular Bragg phenomenon occurs~\cite{J.B.GeddesIII-2006(P)}. There is no such local maximum evident in Figure~\ref{F: Durations}. We attribute this effect to the wide bandwidth of the videopulses as compared to the longer--duration pulses studied earlier. The wide bandwidth effectively swamps the circular Bragg phenomenon.

The center--of--energy speeds increase modestly with increasing wavelength; and they decrease, for given carrier wavelength and polarization state, for the nonlinear case as compared to the linear one. However, the peak and correlation speeds remain roughly constant with increasing carrier wavelength, close in value, and larger than the center--of--energy speeds. And the peak and correlation speeds decrease, for given carrier wavelength and polarization state, from the linear to the nonlinear case, but only slightly. The rate of increase in center--of--energy speed with increasing carrier wavelength is less for videopulses than for the longer--duration pulses we studied earlier~\cite{J.B.GeddesIII-2006(P)}.

We expect identification of these trends to be useful to designers of chiral STF--based devices. Such devices shall be limited to shaping pulses of bandwidth narrower than a certain value, due to the swamping of the circular Bragg phenomenon by shorter--duration pulses.


\bigskip

{\bf Acknowledgments:} J. B. Geddes III gratefully acknowledges the support of an NSF Graduate Fellowship and a SPIE Educational Scholarship.


\begin{thebibliography}{99}

\bibitem{A.Lakhtakia-2005(B)}
A. Lakhtakia and R. Messier,
Sculptured Thin Films: Nanoengineered Morphology and Optics,
SPIE Press, Bellingham, WA, USA (2005). 

\bibitem{J.B.GeddesIII-2006(C)}
J. B. Geddes III, 
Towards shaping of pulsed plane waves in the time domain via chiral sculptured thin films, 
in Frontiers in Optical Technology: Materials and Devices, 
P. K. Choudhury and O. N. Singh, Editors,
Nova Science Publishers, Hauppauge, NY, USA (2006).

\bibitem{J.B.GeddesIII-2002(P)}
J. B. Geddes III and A. Lakhtakia, 
Videopulse bleeding in axially excited chiral sculptured thin films in the Bragg regime, 
Eur. Phys. J. Appl. Phys. 17 (2002) 21--24.

\bibitem{J.B.GeddesIII-2006(P)}
J. B. Geddes III and A. Lakhtakia,
Quantification of optical pulsed--plane--wave--shaping by chiral sculptured thin films,
J. Mod. Opt. (in press).

\bibitem{V.C.Venugopal-1998(P)}
V. C. Venugopal and A. Lakhtakia,
Second harmonic emission from an axially excited slab of a dielectric thin--film helicoidal bianisotropic medium,
Proc. Roy. Soc. Lond. A 454 (1998) 1535--1571.  

\bibitem{R.W.Hellwarth-1977(P)}
R. W. Hellwarth,
Third--Order optical susceptibilities of liquids and solids,
J. Prog. Quant. Electr. 5 (1977) 1--68. 

\bibitem{R.N.Bracewell-2000(B)}
R. N. Bracewell,
The Fourier Transform and Its Applications, 3rd Edition,
McGraw Hill, New York, NY, USA (2000). 

\bibitem{S.C.Bloch-1977(P)}
S. C. Bloch,
Eighth velocity of light, 
Am. J. Phys. 45 (1977) 538--549. 

\bibitem{Y.Jaluria-1996(B)}
Y. Jaluria,
Computer Methods for Engineering,
Taylor \& Francis, Washington, DC, USA (1996).

\bibitem{J.B.GeddesIII-2001(P)}
J. B. Geddes III and A. Lakhtakia,
Reflection and transmission of optical narrow--extent pulses by axially excited chiral sculptured thin films,
Eur. Phys. J. Appl. Phys. 13 (2001) 3--14. Erratum: 16 (2001) 247. 

\bibitem{A.Lakhtakia-1999(P)}
A. Lakhtakia,
Spectral signatures of axially excited slabs of dielectric thin--film helicoidal bianisotropic mediums,
Eur. Phys. J. Appl. Phys. 8 (1999) 129--137. 

\bibitem{J.B.GeddesIII-2003(P)}
J. B. Geddes III and A. Lakhtakia, 
Effects of carrier phase on reflection of optical narrow--extent pulses from axially excited chiral sculptured thin films,
Opt. Commun. 225 (2003) 141--150.

\end{thebibliography}
\end{document}